# Geometry of escort distributions

Sumiyoshi Abe

*Institute of Physics, University of Tsukuba, Ibaraki 305-8571, Japan*

Given an original distribution, its statistical and probabilistic attributes may be scanned by the associated escort distribution introduced by Beck and Schlögl and employed in the formulation of nonextensive statistical mechanics. Here, the geometric structure of the one-parameter family of the escort distributions is studied based on the Kullback-Leibler divergence and the relevant Fisher metric. It is shown that the Fisher metric is given in terms of the generalized bit-variance, which measures fluctuations of the crowding index of a multifractal. The Cramér-Rao inequality leads to the fundamental limit for precision of statistical estimate of the order of the escort distribution. It is also quantitatively discussed how inappropriate it is to use the original distribution instead of the escort distribution for calculating the expectation values of physical quantities in nonextensive statistical mechanics.

# I. INTRODUCTION

The concept of the escort distributions has been introduced by Beck and Schlögl [1] in order to scan the attributes of the original distributions describing the (multi)fractal features of nonlinear dynamical systems. Let $\{p_i\}$ be the original distribution. Then, the escort distribution associated with it is given by [1]

$$P_i = \frac{\phi(p_i)}{\sum_j \phi(p_j)}, \qquad (1)$$

where $\phi$ is a certain positive function. Of particular importance is the case, $\phi(s) = s^q$ ($0 \leq s \leq 1$, $q > 0$), and correspondingly

$$P_i^{(q)} = \frac{(p_i)^q}{\sum_j (p_j)^q}. \qquad (2)$$

The parameter, $q$, is referred to as the order of $P_i^{(q)}$. We mention that the (generalized) expectation value with respect to $P_i^{(q)}$ in eq. (2), termed the $q$-expectation value, plays a crucial role in the formulation of nonextensive statistical mechanics [2,3]. There exists intriguing property [1] concerning the composition law and the group-theoretic structure behind eq. (2). Regarding $p_i \to P_i^{(r)}$ as a transformation, $P_i^{(q)}$ changes as

$$P_i^{(q)} \rightarrow \frac{(p_i)^{qr}}{\sum_j (p_j)^{qr}} = P_i^{(qr)}. \tag{3}$$

Therefore, this transformation forms a one-parameter Abelian group with the identity transformation corresponding to the order unity. Physical significance of this emerging symmetry, however, does not seem to be revealed yet.

In this article, we study the geometric structure of the space of escort distributions. We calculate the Fisher metric along the curve defined by the one-parameter family $\{P_i^{(q)}\}$ and show that it is given by the generalized bit-variance, which describes fluctuations of the crowding index of a multifractal. The Cramér-Rao bound is then discussed in connection with statistical estimate of the order of the escort distribution. We also discuss the Kullback-Leibler divergence between the original and associated escort distributions in the case of the $q$-exponential distribution in nonextensive statistical mechanics, in which the use of the escort distribution for calculating the expectation values is essential. In this way, it is quantified how wrong it is to employ the ordinary definition of the expectation value in nonextensive statistical mechanics.

## II. FISHER METRIC AND GENERALIZED BIT-VARIANCE

Let us start by summarizing the basic issues relevant to our discussion. Consider two distributions, $\{\pi_i\}$ and $\{\pi'_i\}$. The distance between them may be measured by the symmetric Kullback-Leibler divergence

$$D[\pi, \pi'] = K[\pi \| \pi'] + K[\pi' \| \pi]. \tag{4}$$

In this equation, $K[\pi \| \pi']$ stands for the Kullback-Leibler relative entropy [4]

$$K[\pi \| \pi'] = \sum_i \pi_i \ln \frac{\pi_i}{\pi'_i}, \tag{5}$$

which is positive semidefinite and vanishes if and only if $\pi_i = \pi'_i$ ($\forall i$).

Suppose $\pi_i$ be dependent on a set of parameters $\mathbf{q} = (q^1, q^2, \cdots, q^n)$: $\pi_i = \pi_i(\mathbf{q})$. $\mathbf{q}$ supplies a local coordinate in the *n*-dimensional submanifold of the functional space of distributions. Then, the induced metric on this submanifold may be constructed from eq. (4). In fact, taking $\pi'_i$ in the neighborhood of $\pi_i$, that is, $\pi'_i = \pi_i(\mathbf{q} + d\mathbf{q})$, eq. (4) is calculated to yield the first fundamental form

$$ds^2 = D[\pi, \pi'] = \sum_{\mu, \nu = 1}^{n} g_{\mu\nu}(\mathbf{q}) \, dq^\mu \, dq^\nu, \tag{6}$$

where $g_{\mu\nu}$ is given by

$$g_{\mu\nu}(\mathbf{q}) = \sum_i \frac{\partial_\mu \pi_i(\mathbf{q}) \, \partial_\nu \pi_i(\mathbf{q})}{\pi_i(\mathbf{q})} \tag{7}$$

with the notation $\partial_\mu = \partial / \partial q^\mu$. The quantity in eq. (7) is called the Fisher metric, or the Fisher information, in the literature [5]. It defines the Riemannian geometric structure of the submanifold.

Now, we regard the order of the escort distribution, $q$, as a parameter. Accordingly, a curve is defined by the one-parameter family $\{P_i^{(q)}\}$. To study geometry associated with the escort distributions, we measure the distance between $P_i^{(q)}$ and $P_i^{(q+dq)}$ following the line mentioned above. From eq. (2), we obtain

$$\frac{\partial P_i^{(q)}}{\partial q} = \left(<I>_q - I_i\right) P_i^{(q)}, \tag{8}$$

where $I_i$ and $<I>_q$ are the information content and its $q$-expectation value, respectively given by

$$I_i = -\ln p_i, \tag{9}$$

$$<I>_q = \sum_i I_i P_i^{(q)}. \tag{10}$$

We notice that in deriving eq. (8) the original distribution is assumed to be independent of $q$. (The situation becomes different in Sec. III.) Using eq. (7) in the one-dimensional case, we find

$$ds^2 = D[P^{(q)}, P^{(q+dq)}] = (\Delta_q I)^2 \, dq^2. \tag{11}$$

Here, the Fisher metric, $(\Delta_q I)^2$, is the generalized variance [6] of the information content:

$$(\Delta_q I)^2 = <I^2>_q - <I>_q^2. \tag{12}$$

This is a generalization of the so-called bit-variance (i.e., the second bit-cumulant) [1], and is different from the one discussed in Refs. [7-10], in which statistical properties of chaotic dynamical systems are investigated. The ordinary bit-variance is recovered in the limit $q \to 1$. Thus, we find that the bit-variance possesses the geometric meaning as the Fisher metric associated with the escort distributions with the different values of the order.

Now, we wish to point out the relevance of the present discussion to multifractals [1]. Let $l$ be the size of a small box whose collection can cover the multifractal phase space. The probability attributed to the *i*th box (centered at a certain point) reads

$$p_i(l) \sim l^{\alpha_i}, \tag{13}$$

where $\alpha_i$ is the Lipschitz-Hölder exponent, or is commonly referred to as the crowding index [1]. Eq. (13) means that the crowding index is essentially the information content. Therefore, we see that the Fisher metric in eq. (12) provides fluctuations of the crowding index of a multifractal with a natural geometric interpretation.

Finally, we briefly mention parameter estimation theory for the order of the escort distribution. This problem is analogous to estimating temperatures of thermodynamic systems [11]. It is known that the Fisher information is related to the problem of statistical parameter estimation. For the unbiased estimator of $q$, the size of error in estimation, $\delta q$, obeys the Cramér-Rao inequality [5,11]

$$(\delta q)^2 \times (\Delta_q I)^2 \geq 1. \tag{14}$$

Therefore, the generalized bit-variance as the metric gives the fundamental limit for precision of estimate of the order of the escort distribution.

### III. KULLBACK-LEIBLER DIVERGENCE
### FOR $q$-EXPONENTIAL DISTRIBUTION

This section is devoted to calling attention to crucial importance of employing the escort distribution for calculating the expectation values of physical quantities in nonextensive statistical mechanics [2,3,12] based on the Tsallis entropy [13]. This is because there still seems to be confusion caused by published works that employ the ordinary definition of the expectation values with respect to the original distribution.

The stationary distribution in nonextensive statistical mechanics is the "$q$-exponential distribution", which has the following form:

$$\tilde{p}(x) = \frac{1}{Z_q(\lambda)} e_q(-\lambda x), \tag{15}$$

$$Z_q(\lambda) = \int dx\, e_q(-\lambda x), \qquad (16)$$

where $\lambda$ is the factor related to the Lagrange multiplier associated with the $q$-expectation value of $x$ calculated by the use of the escort distribution, as in eq. (10). (In the subsequent discussion, $\lambda$ is nothing but a parameter.) Also, here and hereafter, a continuous random variable, $x\,(\geq 0)$, is considered for the sake of simplicity of the discussion. In eqs. (15) and (16), $e_q(s)$ stands for the "$q$-exponential function" defined by

$$e_q(s) = \left(1 + (1-q)s\right)_+^{1/(1-q)} \qquad (17)$$

with the notation: $(A)_+ \equiv \max\{0, A\}$. In the limit $q \to 1$, eq. (15) tends to the familiar Boltzmann-Gibbs-type exponential distribution. Clearly, the distribution in eq. (15) is normalizable if and only if $q < 2$, leading to $Z_q(\lambda) = [(2-q)\lambda]^{-1}$.

In the case when $1 < q < 2$, the original distribution in eq. (15) is of the Zipf-Mandelbrot type and decays as a power law for large values of $x$. Accordingly, all of its moments are divergent. This is a feature in common with the Lévy distribution in half space, and in fact the original distribution converges to the Lévy distribution by the repeated convolution operations in conformity with the Lévy-Gnedenko generalized central limit theorem [14]. Therefore, in this case, the use of the $q$-expectation values is essential for calculating finite moments to investigate, e.g., the maximum entropy principle. Furthermore, it has been shown [15] that in nonextensive statistical mechanics the ordinary expectation value has to be replaced by the $q$-expectation value, in order to be consistent with the nonextensive generalization of the method of steepest descents developed by Fowler and Darwin for establishing the statistical foundation for the canonical ensemble theory.

The situation is different when $0 < q < 1$. In this case, the original distribution in eq. (15) is support compact, $0 \leq x \leq [(1-q)\lambda]^{-1}$, and therefore all the moments are finite. However, the use of the escort distribution still has to be respected, since eq. (15) itself is derived by maximization of the Tsallis entropy under the constraint on the $q$-expectation value of $x$. (For the entropic basis of the q-exponential distribution, see Ref. [16].) To quantify the discrepancy between the original distribution, $\tilde{p}(x)$, and its associated escort distribution, $\tilde{P}^{(q)}(x) = [\tilde{p}(x)]^q / \int dx' \, [\tilde{p}(x')]^q$, again we calculate the Kullback-Leibler divergences. The results are

$$K[\tilde{P}^{(q)} \| \tilde{p}] = 1 - q - \ln(2-q), \tag{18}$$

$$K[\tilde{p} \| \tilde{P}^{(q)}] = \ln(2-q) - \frac{1-q}{2-q}, \tag{19}$$

$$D[\tilde{P}^{(q)}, \tilde{p}] = \frac{(1-q)^2}{2-q}, \tag{20}$$

which turn out to hold also in the case $1 < q < 2$. In the range $0 < q < 1$, $D$ monotonically decreases from 1/2 to 0 as $q$ varies from 0 to 1. For example, $q = 1/2$ gives rise to $D = 1/6$, which is about 33% of the value of the largest value of the discrepancy, $1/2$.

This fact quantitatively illustrates how wrong it is to employ the original distribution to calculate the expectation values in nonextensive statistical mechanics.

## IV.  CONCLUSION

We have studied geometry associated with the escort distributions by making use of the Kullback-Leibler divergence and the Fisher metric. We have shown that three concepts, i.e., the lower bound of error in estimation of the order of the escort distribution, he bit-variance, and fluctuations of the crowding index of a multifractal, are all endowed with their geometric interpretations. We have also quantitatively discussed how inappropriate it is to use the original distribution instead of the escort distribution in nonextensive statistical mechanics.

In the present work, we have employed the ordinary Kullback-Leibler divergence. There is a possibility of using the generalized Kullback-Leibler divergence (associated with the Tsallis entropy [13]) introduced and discussed in Refs. [17-22]. As shown in Refs. [17,18], in this case, the corresponding Fisher metric is globally conformally equivalent to the ordinary one in eq. (7).